\documentclass[aps,pre,twocolumn,titlepage,nofootinbib]{revtex4}

\usepackage{graphicx}
\usepackage{color}
\usepackage{dcolumn}
\usepackage{latexsym}
\usepackage[normalem]{ulem}
\usepackage{hyperref,amssymb}
\usepackage{url}
\usepackage{color,soul}
\usepackage{graphicx}
\usepackage{verbatim}
\usepackage{multirow}
\usepackage{amsmath}
\newcommand{\beq}{\begin{eqnarray}}
\newcommand{\eeq}{\end{eqnarray}}
\usepackage{mathrsfs}
\usepackage{float,soul}
\usepackage[dvipsnames]{xcolor}
\usepackage{mathtools}
\usepackage{slashed}
\usepackage{graphicx}   
\usepackage{epstopdf}
\usepackage{subfigure}  
\usepackage{hyperref}   
\usepackage{bbold}
\usepackage{wasysym}
\usepackage{feynmp}
\usepackage{hyperref}
\hypersetup{colorlinks,
}

\usepackage[latin1]{inputenc}

\begin{document}

\title{\Large Correlation between optical phonon softening and superconducting $T_c$ in YBa$_2$Cu$_3$O$_x$ within $d$-wave Eliashberg theory}
\author{Cunyuan Jiang$^{1,2,3}$}
\author{Giovanni Alberto Ummarino$^{4}$}
\author{Matteo Baggioli$^{1,2,3}$}
\thanks{b.matteo@sjtu.edu.cn}
\author{Efthymios Liarokapis$^5$}
\author{Alessio Zaccone$^6$}
\thanks{alessio.zaccone@unimi.it }

\address{$^1$ School of Physics and Astronomy, Shanghai Jiao Tong University, Shanghai 200240, China}
\address{$^2$ Wilczek Quantum Center, School of Physics and Astronomy, Shanghai Jiao Tong University, Shanghai 200240, China}
\address{$^3$ Shanghai Research Center for Quantum Sciences, Shanghai 201315,China}
\address{$^4$ Politecnico di Torino, Dipartimento di Scienza Applicata e Tecnologia, corso Duca degli Abruzzi 24, 10129 Torino, Italy.
}
\address{$^5$ Department of Physics, National Technical University
Athens 15780, Greece}
\address{$^6$ Department of Physics ``A. Pontremoli'', University of Milan, via Celoria 16,
20133 Milan, Italy}

\begin{abstract}
We provide a mathematical description, based on d-wave Eliashberg theory, of the strong correlation between the experimentally observed  softening of Raman modes associated with in-plane oxygen motions and the corresponding superconducting critical temperature $T_c$, as a function of oxygen doping $x$, in YBa$_2$Cu$_3$O$_x$. The theoretical model provides a direct link between physical trends of soft optical $A_g$ (in-plane) oxygen modes, the level of oxygen doping $x$, and the superconducting $T_c$. Different regimes observed in the trend of $T_c$ vs doping can be related to corresponding regimes of optical phonon softening in the Raman spectra. These results provide further evidence related to the physical origin of high-temperature superconductivity in rare-earth cuprate oxides and to the significant role of electron-phonon coupling therein.
\end{abstract}

\maketitle
\section{Introduction}
High-temperature cuprates superconductors were discovered by Bednorz and M\"{u}ller \cite{Bednorz1986} under the guiding assumption that strong Jahn-Teller type lattice distortions, akin to those in ferroelectric materials, would lead to large polarons and very strong electron-phonon couplings. Over the subsequent decades much new physics in these materials has been unveiled, where non-trivial magnetic effects accompany the superconducting transition, but a microscopic theory is still lacking \cite{Zhou2021}. 
Several arguments have been put forward supporting an unconventional pairing mechanism in the cuprates, among which the strongest arguments are perhaps the fluctuating antiferromagnetic ordering, suggesting electron-electron interactions \cite{Dahm2009}, and the strongly 2D physics within the layers \cite{Allen2001}. 

In parallel with research on possible unconventional mechanisms, some experimental evidence still points towards a large electron-phonon coupling \cite{Lanzara2001,Lanzara2004,Allen2001} and the isotope effect \cite{Aiura2008,isotoppnas}, possibly enhanced by electronic and magnetic correlations \cite{Louie2021,Shapiro2021}, such that suitably modified Bardeen-Cooper-Schrieffer (BCS) scenarios may be able to account for the observations in doped cuprates, without the need of exotic physics. For an up-to-date discussion of these issues, see \textit{e.g.} Refs. \cite{Schrodi2021,Hirschfeld}.\\

In this work, we add further evidence in favour of an underlying dominant electron-lattice coupling, by proving a description of the superconducting $T_c$ that, crucially, incorporates the optical phonon softening as measured experimentally. In particular, earlier experimental results based on Raman scattering by Liarokapis, M\"{u}ller, Kaldis and co-workers showed that the in-plane $A_g$ oxygen mode displays a dramatic softening upon increasing the oxygen doping $x$ in the model cuprate system YBa$_2$Cu$_3$O$_x$ \cite{Liarokapis1,Liarokapis2}. Concomitantly, the superconducting $T_c$ is experimentally observed to grow in a correlated fashion with the increase of doping $x$ up to a maximum or dome. This correlation has remained thus far as a mere observational experimental fact, in spite of the crucial importance of oxygen doping for high-T superconductors performance \cite{Poccia2011}.
Furthermore, very recent results based on X-ray absorption spectra support a striking correlation between anharmonic lattice displacements of Sr atoms (via their coupling to apical oxygens) and the superconducting critical temperature in  YSr$_2$Cu$_{2.75}$Mo$_{0.25}$O$_{7.54}$ and Sr$_2$CuO$_{3.3}$ \cite{Conradson_2024}. Again, this kind of correlations cannot be explained within unconventional (\textit{e.g.} spin fluctuations or electron repulsion) mechanisms.

We build on the recent observation that sharp optical (or acoustic) phonon softening implemented within Migdal-Eliashberg theory can lead to huge electron-phonon couplings and consequent order-of-magnitude increase in $T_c$ \cite{Jiang_2023}. This generic theoretical prediction has been recently verified by means of first-principles computations on the example of a new compound, of Zr$_{2}$S$_{2}$C \cite{Jamwal_2024}. It was observed that, as predicted by the theory \cite{Jiang_2023}, increasing the depth of the acoustic phonon softening can increase the electron-phonon coupling $\lambda$ by a factor 2 \cite{Jamwal_2024}. 

Implementing the experimentally observed softening of the $A_g$ mode as a function of doping $x$ within the d-wave Migdal-Eliashberg scheme, we mathematically describe the correlation between the oxygen doping, the increase of the softening depth (and sharpening of its width) and the concomitant non-trivial variation of the $T_c$ with doping, in good qualitative agreement with experimental data. In the comparison between theory and experiments, three free fitting parameters are used: one is the electron-phonon matrix element $C$ which controls the quantitative value of $T_c$. The other fitting parameter is the shape of the optical phonon softening. Both parameters vary with $x$. The evolution of the softening with the doping level $x$ is found to be physically meaningful and in qualitative agreement with typical momentum-resolved experimental observations of optical softening in the cuprates \cite{PhysRevB.106.155109}. The dependence of the electron-phonon matrix elements on the doping $x$ will be important to capture the correct order of magnitude of $T_c$ and will be discussed at length in the following. Finally, the third free parameter is the softening wave-vector, which, as we will see, does not affect significantly the results and will be therefore taken in a first approximation as doping-independent (see below for a more detailed discussion on this point).

Numerous experimental studies have established that the plateau in the $T_c - x$ (oxygen doping) phase diagram corresponds to the suppression of $T_c$ induced by charge order, charge density wave (CDW) and charge density fluctuations (CDF) \cite{cdw1}. Also, there are many works confirming that the CDW can give rise to phonon softening at a CDW wavevector $q_{\text{CDW}}$ \cite{chaix2017dispersive,cdw2}. For YBCO systems, it has been confirmed that the intensity of CDF reaches its maximum at near optimal doping $x=7$, where the phonon is softened to the lowest frequency \cite{arpaia2023signature}.
There is a strong correlation between the in-phase oxygen mode frequency and the characteristic doping levels of YBa$_2$Cu$_3$O$_x$, shown in \cite{LiarokapisPRL}. This correlation was discussed in terms of the superstructures developed from the chain oxygen atoms, but it is, in fact, related to the charge ordering.
As a matter of fact, a description of these softened-phonon effects on the $T_c$, within the conventional phonon mediated superconducting framework, is still missing. The simultaneous interplay between charge order, anharmonic phonons and superconductivity still poses an enormous challenge to microscopic theories \cite{Sarkar,Setty_2024,kagome_2024}. Both older theoretical and experimental works \cite{isotoppnas,Zeyher1990}, as well as very recent experimental evidence \cite{Conradson_2024}, suggest that phonons cannot be overlooked for explaining the superconductivity in doped cuprates. Our results provide further evidence to the role of electron-phonon coupling for the physical origin of high-temperature superconductivity in oxygen doped rare-earth cuprate oxides.

\section{Strategy}
We choose an Occam's razor strategy and work with a minimal mathematical model in order to: (i) reduce the number of unknown adjustable parameters and (ii) unveil direct connections between experimentally observable trends.
Also, working with an effective model allows us to accommodate the experimental input from Raman scattering experiments, which would not be possible within a fully ab-initio approach. In turn, this allows us to focus on physical trends rather than on absolute values of physical quantities, such as $T_c$, which are anyway out of reach even for the most detailed microscopic calculations. 
It should be noted that the Raman scattering data that we used \cite{Liarokapis1,Liarokapis2,Liarokapis2019} were obtained at room temperature, for many values of oxygen content, several
(7-8) micro-crystallites and long acquisition times to secure good
statistics. Data obtained at low liquid nitrogen temperatures ($T=77$ K) show the
same characteristics as the room temperature ones, though for high
oxygen concentrations the data were collected below $T_c$.
We show also the Raman measurements taken at low temperature ($T=77$ K) in the Appendix. Since these data present exactly the same features of phonon softening observed at room temperatures, we use the room temperature data as input for the theoretical model, given their better statistics.

\begin{figure}
    \centering
    \includegraphics[width=0.8\linewidth]{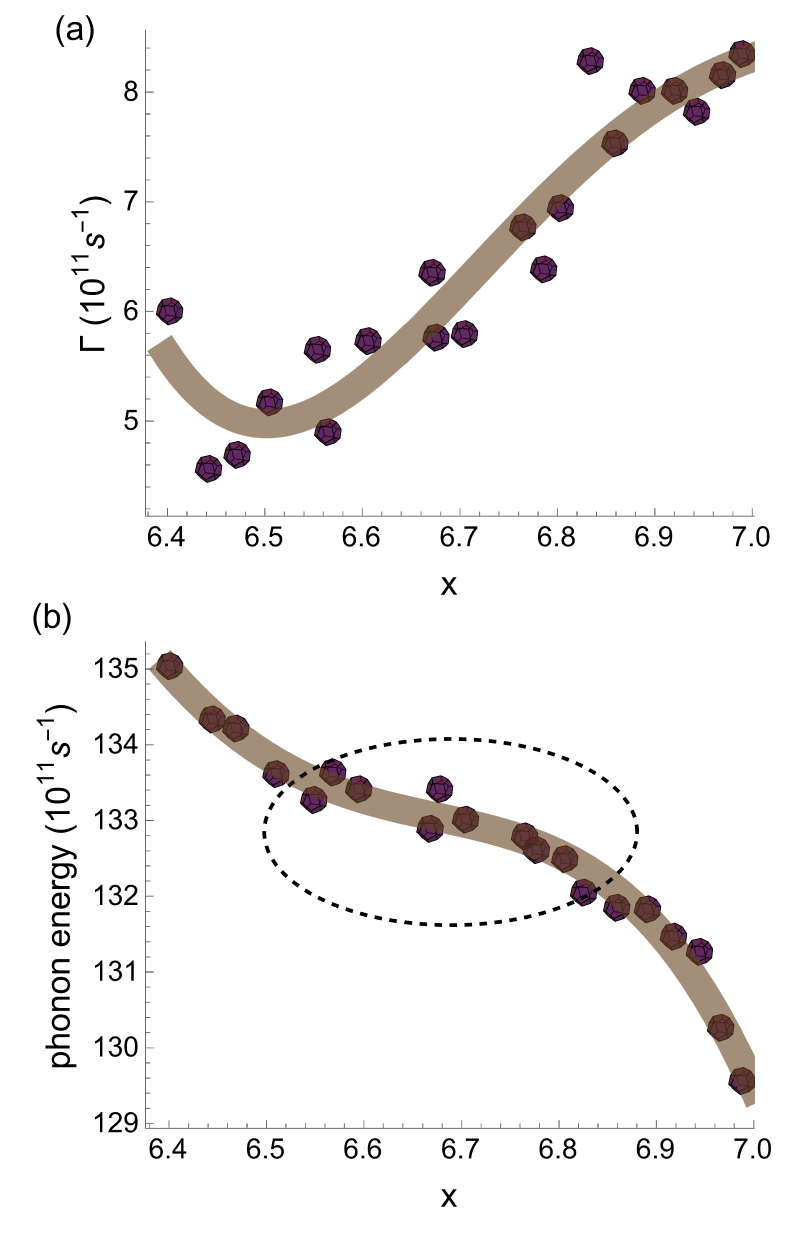}
    \caption{\textbf{(a)} The experimental data (dark symbols) for the linewidth of the $A_g$ optical phonon as a function of the doping level $x$ reproduced from \cite{Liarokapis2017}. \textbf{(b)} The experimental data (dark symbols) for the $A_g$ optical phonon energy reproduced from \cite{Liarokapis1,Liarokapis2019} as well. Both measurements are performed at small wave-vector, near the $\Gamma$ point. The thick brown lines are the phenomenological fitting functions used as experimental input into the theoretical model. In particular, the phonon linewidth is associated to the parameter $\gamma$ defined in Eq.\eqref{line} and the phonon energy with the energy at the softening point which corresponds to $h(1-\delta)$ in Eq.\eqref{dispersionmodel}. The dashed circle emphasize the plateau feature which appears in the phonon energy at intermediate doping levels.}
    \label{fig:1}
\end{figure}

The strategy behind our theoretical modeling is as follows:

(i) The CDW softens the phonon dispersion relation at $q_{\text{soft}}=q_{CDW}$. There are three parameters controlling the properties (shape) of the softened dispersion relation: the depth of softening $\delta$, which can be determined from experimental data as shown in Fig.\ref{fig:1} (b), the width of the softening region, $\beta$, which will be discussed more in detail as an adjustable parameter, and the softening wave vector $q_{\text{soft}}$ which is, usually, fixed. The linewidth of the phonon is also taken directly from experimental data as shown in Fig.\ref{fig:1} (a). The effective dispersion relation and the effect of depth, width, and position of the softened dispersion relation can be seen in the inset of Fig.\ref{fig:2} (a).

(ii) Once that the softened phonon dispersion relation has been determined, the Eliashberg function $\alpha^2 F(\omega)$ can be obtained by using the method described in Ref. \cite{Jiang_2023}.

(iii) By implementing the Eliashberg function $\alpha^2 F(\omega)$ into the simplified Eliashberg equation for the electron-phonon coupling $\lambda$ \cite{PhysRevB.12.905}, $T_c$ can be obtained via the Allen-Dynes formula, in an approximate way.

(iv) By direct numerical solution of the d-wave Eliashberg equations, we can verify the $T_{c}$ trend versus doping $x$ obtained in the point (iii) and produce a physical interpretation of the results.

In summary, our strategy will be to build an effective phonon-softening dispersion relation with input from experimental data, and to use this effective dispersion relation to compute the superconducting $T_c$ within the conventional phonon-mediated framework. The charge order results in phonon softening at $q_{\text{soft}}=q_{CDW}$, hence, the charge order is implemented indirectly through the dispersion relation for the softened phonon. Other effects of charge order will not be taken into account explicitly to avoid increasing the number of parameters in the model in the absence of suitable experimental input, although this extension may be considered in future work, \textit{e.g.}, along the lines of \cite{Sarkar}. In the following sections, the computational methods will be described with more details.

\section{s-wave Eliashberg calculation}
\subsection{Theoretical method and computational details}
Let us denote with the symbol $q$ the wave-vector and with the symbol $x$ the doping level. We start by considering an optical phonon mode whose dispersion relation $\omega(q,x)$ is given by:
\begin{equation}
    \omega^2-\Omega(q,x)^2+i \omega \Gamma(q,x)=0\,.\label{dispersion}
\end{equation}
Here, $\Omega(q,x)$ is the renormalized energy of the phonon mode and $\Gamma(q,x)$ the corresponding linewidth. We emphasize that the energy $\Omega(q,x)$ contains already the corrections from the real part of the self-energy arising from phonon-phonon scattering, \textit{i.e.}, anharmonicity, and it is not the bare phonon energy. On the other hand, the linewidth $\Gamma(q,x)$ is related to the imaginary part of the self-energy. The spectral function for this mode is given by the standard Lorentzian form
\begin{equation}
    \mathcal{B}(q,\omega,x)=\frac{\omega\Gamma(q,x)}{\pi[(\omega^2-\Omega^2(q,x))^2+\omega^2\Gamma^2(q,x)]}.\label{spec}
\end{equation}
By assuming a constant electronic density of states (DOS) at the Fermi momentum and wave-vector independent electron-phonon matrix elements, the Eliashberg function is given by  (see the detailed derivation in Ref. \cite{Jiang_2023})
\begin{equation}
    \alpha^2 F(\omega,x)=C(x)\int_0^4 \mathcal{B}\left(k_F \sqrt{\zeta},\omega,x\right) d\zeta, \label{optical result}
\end{equation}
where $\zeta=2 k_{F}^{2}(1+\cos \theta)$, with $\theta$ the polar angle of the Fermi sphere, $k_F$ the Fermi momentum, and $C(x)$ is an effective parameter incorporating the microscopic information about the screening and the electron-phonon matrix elements. While the electron-phonon matrix in the cuprates is arguably anisotropic, here $C(x)$ is intended to represent an angular average over the different directions \cite{Carbone}.
Finally, the Fermi wave-vector is fixed to $k_F=0.5 q_{\text{max}}$ (see below for the definition of $q_{\text{max}}$) to make sure that the $\zeta$ variable spans the whole reciprocal space,  $\zeta\in [0,4]$. We can then define the electron-phonon coupling parameter $\lambda$ as customary in Eliashberg's theory \cite{Carbotte2003}
\begin{equation}\label{lala}
\lambda(x)=2 \int_{0}^{\infty}\frac{ \alpha^2 F(\omega,x)}{\omega}d\omega\,.
\end{equation}
Finally, we will compute the critical temperature using the Allen-Dynes formula \cite{PhysRevB.12.905}:
\begin{widetext}
\begin{equation}
      T_c(x)\,=\,\frac{f_1(x)\,f_2(x)\,\omega_{log}(x)}{1.2}\,\exp\left(-\frac{1.04\,(1+\lambda(x))}{\lambda(x)-\mu^\star\,-\,0.62\,\lambda(x)\,\mu^\star}\right),\label{allenformula}
\end{equation}
\end{widetext}
where the logarithmic average frequency of the spectrum is defined as:
\begin{equation}
\omega_{log}(x)=\exp \left(\frac{2}{\lambda(x)}\int_{0}^{\infty} d\omega \frac{\alpha^{2}F(\omega,x)}{\omega} \ln \omega \right).
\end{equation}
Following \cite{PhysRevB.12.905}, we have also defined
\begin{align}
    & f_1(x)=[1+(\lambda(x)/\Lambda_1)^{3/2}]^{1/3},\nonumber\\
    & f_2(x)=1+\dfrac{(\Bar{\omega}_2(x)/\omega_{log}(x)-1)\lambda^2(x)}{\lambda^2(x)+\Lambda_2^2(x)},\nonumber\\
    & \Bar{\omega_2}(x)=\left[\dfrac{2}{\lambda(x)}\int_0^\infty \alpha^{2}F(\omega,x) \omega\, d\omega \right]^{1/2},\nonumber\\
    & \Lambda_1=2.46(1+3.8 \mu^\star),\nonumber\\
    & \Lambda_2(x)=1.82(1+6.3 \mu^\star)(\Bar{\omega}_2(x)/\omega_{log}(x)),\nonumber\\
    & \mu^\star=0.
\end{align}
Here, $\mu^\star=0$ is chosen as typical value for strong-coupled superconductors 
\cite{Dwave4}.

At the heart of the model lies the effective description of the soft optical phonon, in this case the $A_g$ in-plane oxygen mode which is experimentally known to exhibit a large softening upon increasing the doping $x$ \cite{Liarokapis1,Liarokapis2}. We adopt the effective description of Ref. \cite{Jiang_2023} where the softened optical phonon is given in terms of a flat dispersion with a Gaussian-shaped dip:
\begin{align}
    & \Gamma(q,x)=\gamma(x)\,,\label{line}\\
    &\Omega(q,x)=h \left(1-\delta(x)\, e^{-\left({q/q_{\text{max}}}- q_{\text{soft}}/q_{\text{max}}\right)^2/\beta^2(x)}\right). \label{dispersionmodel}
\end{align}
We notice that, within this description, the softening of the optical phonon is parameterized at a specific wave-vector $q_{\text{soft}}$. In other words, the experimental values reported in Fig.\ref{fig:1} (b) for the phonon energy will be identified with $\Omega(q=q_{\text{soft}},x)=h(1-\delta(x))$. On the other hand, the corresponding linewidth, shown in Fig.\ref{fig:1} (a), will be identified with the wave-vector independent quantity $\gamma(x)$. Finally, the constant $h$, which is doping level and wave-vector independent, corresponds to the frequency of the non-softened optical phonon measured experimentally, that is, $h=135.6 \,\cdot 10^{11} \,s^{-1}$.

The physical parameter $\delta(x) \in [0,1]$ is the dimensionless depth of the softening, which can be obtained by fitting the experimental data as shown in Fig.\ref{fig:1} (b). Meanwhile, $\beta(x)$ is a free parameter corresponding to the dimensionless width of the softening in units of the maximum wave-vector, for which we keep it a constant $\beta(x)=0.2$ in our calculation. Moreover, $q_{\text{soft}}$ is the softening wave-vector in units of the maximum wave-vector value $q_{\text{max}}$, which also represents a free parameter in the theoretical model. As explicitly demonstrated below, the choice of its value does not affect qualitatively the predictions of the model. Finally, $q_{\text{max}}$ is used to set the characteristic length scale of the system and its value is therefore unimportant for the analysis below. Physically, it can be thought of as the size of the Brillouin zone.\\

The Raman measurements of the in-phase $A_g$ mode at varying doping level $x$ \cite{Liarokapis1,Liarokapis2019} thus fix the values of $\mathrm{Re}[\omega(q_{\text{soft}})]$ upon varying $x$. This, in turn, fixes $h$ and $\delta$ as a function $x$. Let us emphasize that since $\Gamma \ll \Omega$ (see Fig.\ref{fig:1}), then we can safely assume $\mathrm{Re}[\omega(q_{\text{soft}})]=\Omega(q_{\text{soft}})$ to simplify our analysis. This leads immediately to mapping the Raman experimental data for the phonon energy onto the combination $h(1-\delta(x))$ in our mathematical model, where all the doping dependence goes into the parameter $\delta(x)$. We are thus left with a free parameter $\beta(x)$ corresponding to the width of the softening, which is the only non-trivial free parameter not constrained by the experimental data, along with the overall re-scaling factor $C(x)$ in Eq.\eqref{optical result}. 

\begin{figure}[ht]
    \centering
    \includegraphics[width=\linewidth]{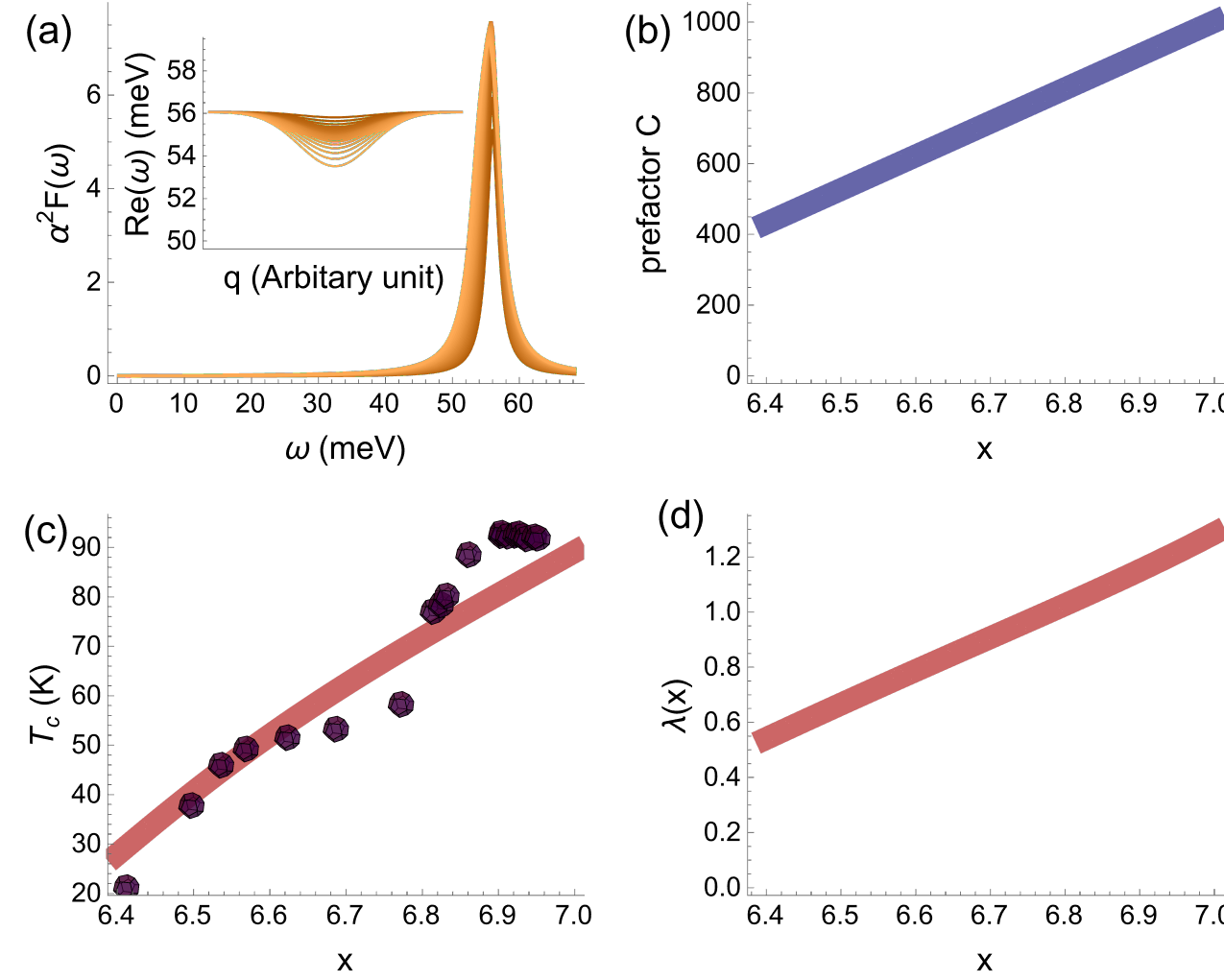}
    \caption{\textbf{(a)}: The inset shows the dispersion relation of the $A_g$ optical phonon mode used in the theoretical model for different values of the doping level $x$. The softening energy is an experimental input from Fig.\ref{fig:1}(b) while the softening width $\beta(x)$ is phenomenologically taken to be a constant, $\beta(x)=0.2$. The softening wave-vector is set at half of the Brillouin zone. Panel (a) shows the corresponding Eliashberg function for different doping levels $x$. From darker to lighter, $x$ increases. \textbf{(b)}: The linear dependence of $C(x)=1013.3x-6053.3$ on the oxygen doping $x$ that produces the theoretical $T_c$ curve in the panel (c). \textbf{(c)}: The theoretically predicted critical temperature (thick curves) together with the experimental data from \cite{Bagley} (symbols). \textbf{(d)}: The electron-phonon coupling $\lambda(x)$ as a function of $x$. In the calculation, $\mu^* =0$ is used throughout.}
    \label{fig:2}
\end{figure}

To reproduce the experimental $T_c$, it is necessary that the value of $C(x)$ depends on the values of doping $x$, as well as $\beta(x)$, and that it displays a physically-reasonable trend compatible with typical momentum-resolved measurements of \cite{PhysRevB.106.155109}. In particular, it is expected that the width of the softening increases when the softening depth decreases \cite{PhysRevB.106.155109,Cohen_softening}. Let us also emphasize that we assume for $C(x)$ a linear dependence from $x$ as the simplest functional dependence that allows one to reproduce, in an approximate way, the experimental $T_c$. This procedure gives the linear dependence shown in Fig.~\ref{fig:2}. Finally, the value of the softened optical phonon linewidth $\Gamma(q,x)$ is set equal to $\gamma(x)$, independent of $q$ as per Klemens' approximation for optical phonon damping \cite{Klemens}, and its value is fixed by the experimental Raman measurements \cite{Liarokapis1,Liarokapis2019} (see panel (b) in Fig.\ref{fig:1}).\\

\subsection{Results}
Since the Raman measurement is performed near the $\Gamma$-point but does not specify the precise value of $q$, we consider two different scenarios for the location of the softening. In the first scenario, we assume the softening to be located in the middle of the Brillouin zone, while in the second scenario we consider the softening right at the $\Gamma$-point, i.e. $q_{\text{soft}}=0$.

\subsubsection{Softening at finite $q$}
We thus proceed to an empirical fitting (interpolation) of the softened $A_g$ mode data of frequency and linewidth from experimental data \cite{Liarokapis1,Liarokapis2019}, as shown in Fig. \ref{fig:1}, as a function of doping level $x$. This will completely determine the form of $\gamma(x)$ and $\delta(x)$ entering the theoretical model. We are then left only with $C(x)$ (a linear function of $x$), $q_{\text{soft}}$ and $\beta(x)$ (allowed to vary with $x$) as fitting parameters, to match the experimentally measured $T_c$ vs $x$ curve. The linear dependence of $C(x)$ from $x$ is fixed also by imposing that the predicted and experimental $T_c$ values agree for the lowest doping $x\approx 6.4$. By fixing $ q_{\textit{soft}}/q_{max}=0.5$ (half of the maximum wave-vector), we obtain the dispersion relation of the phonon mode for different doping levels, which is shown in Fig.\ref{fig:2}(a) and the corresponding Eliashberg function which is obtained using the phenomenological parameter $\beta(x)=0.2$. We have explicitly verified that a variation in $q_{\text{soft}}/q_{max}$ does not strongly affect the final results (see below). All in all, this whole procedure leads to the predicted $T_c$ vs doping $x$ trend shown in Fig. \ref{fig:2}(c). The corresponding electron phonon coupling $\lambda(x)$ as a function of the doping level $x$ is shown in \ref{fig:2}(d).

\begin{figure}
    \centering
    \includegraphics[width=\linewidth]{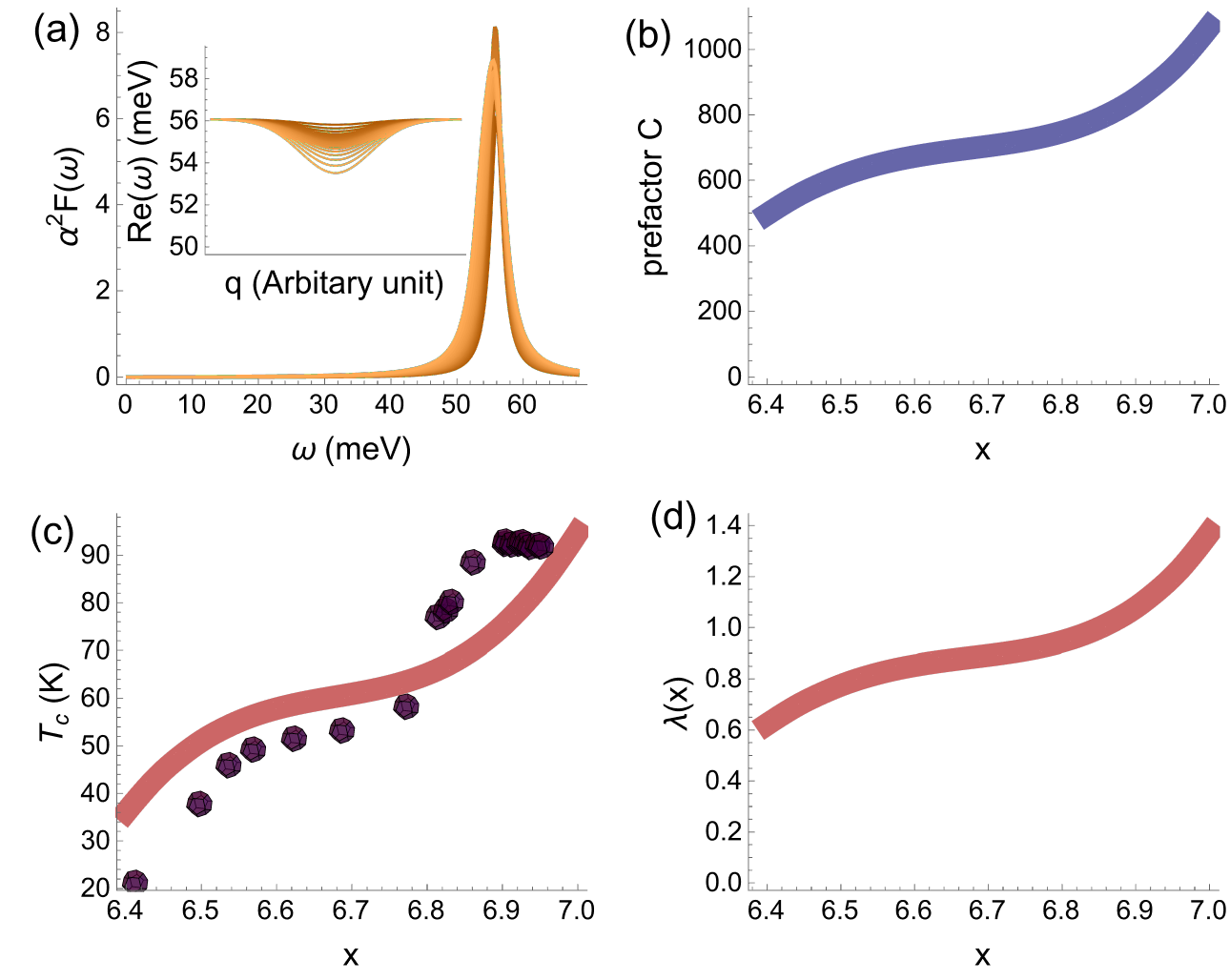}
    \caption{The same analysis as in Fig.\ref{fig:2} but assuming $C(x)=432+14080\delta(x)$ to be softening dependent.}
    \label{fig:3}
\end{figure}

Based on the results shown in Fig.\ref{fig:2}, it can be seen that both the electron-phonon coupling and $T_c$ are dominated by the effective electron-phonon matrix element $C(x)$, while the effect of other parameters like $q_{\text{soft}}$, $\beta(x)$, $\delta(x)$ are weaker. We then attempt to implement $C(x)$ to be softening-dependent, \textit{i.e.} $C(x) \propto \delta(x)$. The results are shown in Fig.\ref{fig:3}, where it can be seen that the plateau of $T_c$ in the range $x=6.6-6.8$ can be qualitatively reproduced.

The above analysis provides an interesting link between in-phase oxygen optical mode softening due to doping and the high-temperature $T_c$ in the slightly overdoped/nearly optimally doped cuprates. In particular, it suggests that the (much debated) flattening of $T_c$ upon increasing doping in the regime $x=6.5-6.8$ is directly linked to the slower decrease of phonon energy with $x$ in that same doping regime: \textit{cfr.} the horizontal oval shaded regions in Fig. \ref{fig:3}(b) and Fig. \ref{fig:1}(b). Less trivial, and even more interesting, is the explanation of the further peak in $T_c$ vs $x$ in the region $x=6.8-7.0$. This effect is directly linked to the sharpening of the softening dip as reflected by the decrease of the width parameter $\beta(x)$, as is evident by comparing the vertical shaded ovals in Fig. \ref{fig:3}(b).\\

\subsubsection{Softening at the $\Gamma$-point, $q_{\text{soft}}=0$}
\begin{figure}
    \centering
    \includegraphics[width=\linewidth]{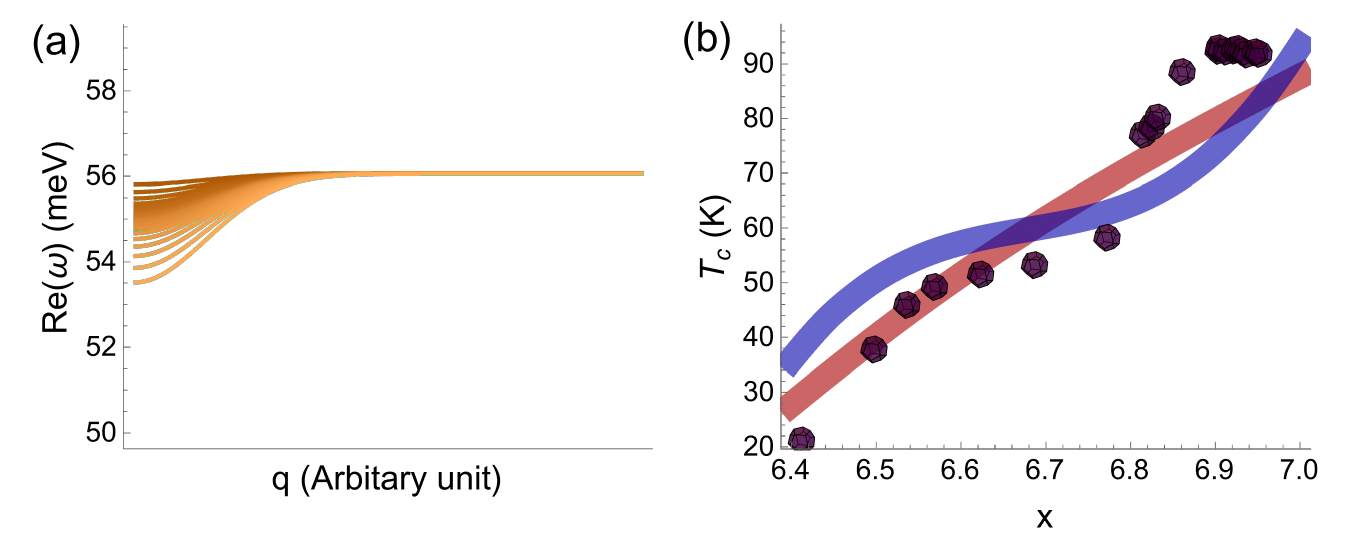}
    \caption{\textbf{(a)} The dispersion relation of the softened optical mode by assuming $q_{\text{soft}}=0$. \textbf{(b)} The corresponding theoretical predictions for $T_c$ using a linear $C(x)$ as in Fig.\ref{fig:2} (red) and a softening dependent $C(x)$ as in Fig.\ref{fig:3} (blue). The black bullets are the experimental data.}
    \label{figqzero}
\end{figure}

As emphasized above, the theoretical model contains, by construction, three unknown physical parameters $C(x),\beta(x), q_{\text{soft}}$ which cannot be determined from the experimental data. The first two have been discussed in detail above. The third parameter, which describes the position of the softening within the Brillouin zone, needs further clarifications. In particular, in our first attempt, we have imposed the phonon softening to occur somewhere between the $\Gamma$-point and the Brillouin zone boundary. Nevertheless, the experimental data from Ref. \cite{Liarokapis1,Liarokapis2019} reported in Fig.\ref{fig:1} are measured at low wave-vector. In order to prove that the choice of $q_{\text{soft}}$ does not affect our results qualitatively, we have performed the same computation by assuming that the softening region appears right at the $\Gamma$ point in the Brillouin zone, coinciding with $q_{\text{soft}}=0$. The corresponding dispersion relation for the softened phonon is shown in Fig.\ref{figqzero} together with the theoretical predictions for $T_c$ (compared to the experimental data) obtained with the two schemes of $C(x)$ linear in $x$ as in Fig.\ref{fig:2} and $C(x)\propto \delta(x)$ as in Fig.\ref{fig:3}. Evidently, the results remain qualitatively the same, proving the subleading effect of the choice of $q_{\text{soft}}$, as already advertised above.\\

\subsection{Physical justification for the $x$-dependence of $C(x)$}
We have seen that the experimental critical temperature can be reproduced, as a first approximation, with $C(x)$ as a linear function of $x$. A reasonable physical justification which supports that $C(x)$ grows linearly with the oxygen doping $x$ goes as follows.
In Ref. \cite{Jiang_2023} the effective parameter $C$ is defined as directly proportional to the electron-phonon matrix element $g_{\boldsymbol{k}\boldsymbol{k'}}$, that is, $C \sim g_{\boldsymbol{k}\boldsymbol{k'}}$. The latter, according to Bardeen-Pines theory accounting for Thomas-Fermi screening, reads as:
\begin{equation}   g_{\boldsymbol{k}\boldsymbol{k'}}^2=|v_{q}^{Z}/\epsilon(0,q)|^2=\left|\frac{4\pi Ze^2}{q}\left(\frac{N}{M}\right)^{1/2}\frac{q^2}{q^2+k_{s}^{2}}\right|^2, \label{gq}
\end{equation}
where $\epsilon(0,q)=1+\frac{k_{s}^{2}}{q^{2}}$ is the static dielectric constant, computed by solving the electrostatic Poisson-equation problem  for an electron gas in presence of the positive background charge of the lattice \cite{Kittel}, and $k_s$ is the screening wave-vector. For a quantum Fermi gas at $T=0$, the screening wave-vector is given by the Thomas-Fermi wave-vector as \cite{Ashcroft}: $k_{TF}^{2}=4 \left(\frac{3 n}{\pi}\right)^{1/3}$,
which is the squared inverse Thomas-Fermi screening length, with $n$ the number density of electrons. At higher temperatures, for a non-degenerate electron gas (the Debye-H{\"u}ckel screening), $k_{s}=\sqrt{\frac{4 \pi  e^2 n}{k_{B}T}}$, with $e$ the electron charge and $k_{B}$ the Boltzmann constant. 
For high-temperature SCs such as YBCO we can assume that we are in the non-degenerate Debye-H{\"u}ckel regime. Hence, upon inserting the Debye-H{\"u}ckel screening $k_s$ into Eq. \eqref{gq}, we obtain, in the regime $k_s^2 \gg q^2$ (low phonon wavevectors),
\begin{equation}
C \sim g_{\boldsymbol{k}\boldsymbol{k'}} \sim 1/n.\label{intermediate}
\end{equation}
At chemical equilibrium, the concentration of electrons, $n$, and the concentration of holes, $p$, are related via \cite{Kittel}:
\begin{equation}
    n\,p=\text{const}(T).
\end{equation}
Hence, at a fixed temperature, $n \sim 1/p$. Since in YBCO the concentration of holes increases proportionally to the oxygen doping $x$ \cite{Talantsev}, upon inserting back into Eq. \eqref{intermediate}, we finally obtain
\begin{equation}
    C(x) \propto x,
\end{equation}
which thus justifies the ansatz used above in our model.

\section{d-wave Eliashberg theory}
In the above approach we neglected the well-document d-wave character of Cooper pairs in the cuprates \cite{Anderson}. In what follows, we implement the phonon softening model and the doping-dependent effective electron-phonon coupling $C(x)$ into the d-wave formulation of Eliashberg theory.
We wish to reproduce, in a approximate way, the experimental critical temperatures versus oxygen doping with a minimal model so we neglect a lot of previous parameters as $q_{\text{soft}}$, $\beta(x)$, $\delta(x)$ and we use just the experimental data: the phonon representative energy $\Omega_{0}(x)$ and the Raman linewith $\Gamma(x)$. We do this only to simplify the calculation as much as possible and to minimize the number of free parameters. In the end we will be able to reproduce the trend of $T_{c}$ as a function of $x$ quite well using a single free parameter for which we will give a physical justification. The first step is to determine the input parameters of the theory as a function of the oxygen doping $x$. We fit the experimental data presented in Fig. \ref{fig:1} with a third-degree polynomial for the Raman frequency,
$\Omega_{0}(x)=8496.79564-3799.76085x+570.27049x^{2}-28.537x^{3}$, and for the linewidth, $\Gamma(x)=6894.76021-3072.10979x+455.86155x^{2}-22.5184x^{3}$. The fit is equally good and indistinguishable from the fifth-order polynomial fit shown in Fig. \ref{fig:1}.
In this way, we can have the input parameters for calculating, in an analytical way, the spectral function $\alpha^{2}F_{s}(\Omega,x)$. 

In what follows, we denote the peak phonon energy with $\Omega_0$, the generic phonon energy with $\Omega$, and the electron energy with $\omega$.

The electron-phonon spectral function now is
\begin{equation}
\begin{split}
\alpha^{2}F_{s}(\Omega,x)=\frac{C(x)\Omega\Gamma(x)}{\pi[(\Omega^{2}-\Omega_{0}(x)^{2})^{2}+\Omega^{2}\Gamma^{2}(x)]}\\=\frac{\lambda_{s}(x)\Omega\Omega_{0}(x)\Gamma(x)}{\pi[(\Omega^{2}-\Omega_{0}(x)^{2})^{2}+\Omega^{2}\Gamma^{2}(x)]}
\end{split}
\end{equation}
where $C(x)=\lambda_{s}(x)\Omega_{0}(x)$ and $\lambda_{s}(x)$ is the value of the electron-phonon coupling constant.
We calculated the critical temperatures by solving, in a numerical way, the one band d-wave Eliashberg equations \cite{Dwave1,Dwave2,Dwave4,Dwave5,Dwave6,Dwave7,Dwave0,Dwave8}. In this case two coupled equations for the gap $\Delta(i\omega_{n},\phi)$ and renormalization functions $Z(i\omega_{n},\phi)$ have to be solved ($\omega_{n}$ denotes the Matsubara frequencies and $\phi$ and $\phi'$ are the azimuthal angles of $k$ and $k'$ in the $ab$ plane). The d-wave one-band Eliashberg equations in the imaginary axis representation, when the Migdal theorem is valid \cite{Migdal}, are:
\begin{align}
\omega_{n}Z(\omega_{n},\phi)=&\omega_{n}+\nonumber \\ & \pi T
\sum_{m}\int_{0}^{2\pi}\frac{d\phi'}{2\pi}\Lambda(\omega_{n},\omega_{m},\phi,\phi')N_{Z}(\omega_{m},\phi')
\end{align}
\begin{equation}
\begin{split}
Z(\omega_{n},\phi)\Delta(\omega_{n},\phi)=\pi T
\sum_{m}\int_{0}^{2\pi}\frac{d\phi'}{2\pi}[\Lambda(\omega_{n},\omega_{m},\phi,\phi')-\\
\mu^{*}(\phi,\phi')\Theta(\omega_{c}-|\omega_{m}|)]N_{\Delta}(\omega_{m},\phi')
\label{eq:EE2}
\end{split}
\end{equation}
where $\Theta(\omega_{c}-|\omega_{m}|)$ is the Heaviside function, $\omega_{c}$ is a
cut-off energy and
\begin{equation}
\begin{split}
&\Lambda(\omega_{n},\omega_{m},\phi,\phi')=\\
&=2\int_{0}^{+\infty}\Omega d\Omega
\alpha^{2}F(\Omega,\phi,\phi')/[(\omega_{n}-\omega_{m})^{2}+\Omega^{2}],
\end{split}
\end{equation}
\begin{equation}
N_{Z}(\omega_{m},\phi)=
\frac{\omega_{m}}{\sqrt{\omega^{2}_{m}+\Delta(\omega_{m},\phi)^{2}}},
\end{equation}
\begin{equation}
N_{\Delta}(\omega_{m},\phi)=
\frac{\Delta(\omega_{m},\phi)}{\sqrt{\omega^{2}_{m}+\Delta(\omega_{m},\phi)^{2}}}.
\end{equation}
We assume \cite{Dwave1,Dwave2,Dwave4,Dwave5,Dwave6,Dwave7} that the electron-phonon spectral function $\alpha^{2}(\Omega)F(\Omega,\phi,\phi')$ and the Coulomb
pseudopotential $\mu^{*}(\phi,\phi')$ at the lowest order contain
separated s- and d-wave contributions $\alpha^{2}(\Omega)F_{s,d}(\Omega,x)$ and $\mu^{*}_{s,d}(\omega_{c})$,
\begin{equation}
    \begin{split}
\alpha^{2}F(\Omega,\phi,\phi')&=\lambda_{s}\alpha^{2}F_{s}(\Omega)
\\
&+\lambda_{d}\alpha^{2}F_{d}(\Omega)\sqrt{2}\cos(2\phi)\sqrt{2}\cos(2\phi'),
\end{split}
\end{equation}
\begin{equation}
\mu^{*}(\phi,\phi')=\mu^{*}_{s}(\omega_{c})
+\mu^{*}_{d}(\omega_{c})\sqrt{2}\cos(2\phi)\sqrt{2}\cos(2\phi')
\end{equation}
as well as the self energy functions:
\begin{equation}
Z(\omega_{n},\phi)=Z_{s}(\omega_{n})+Z_{d}(\omega_{n})\cos(2\phi),
\end{equation}
\begin{equation}
\Delta(\omega_{n},\phi)=\Delta_{s}(\omega_{n})+\Delta_{d}(\omega_{n})\cos(2\phi).
\end{equation}
The spectral functions $\alpha^{2}F_{s,d}(\Omega)$ are normalized in a way that $2\int_{0}^{+\infty}\frac{\alpha^{2}F_{s,d}(\Omega)}{\Omega}d\Omega=\lambda_{s,d}$
and, we choose, for simplicity, $\lambda_{s}=\lambda_{d}$. Also, we assume that $\alpha^{2}F_{s}(\Omega)=\alpha^{2}F_{d}(\Omega)$ i.e. also the shape of the spectral functions are the same. In this case we search for solutions of the Eliashberg equations in a pure d-wave form, as indicated by the experimental data, for the gap function $\Delta(\omega,\phi')=\Delta_{d}(\omega)\cos(2\phi)$ (the $s$ component is zero and this happens for example when \cite{Dwave4} $\mu^{*}_{s} \gg \mu^{*}_{d}$). In the more general case $\Delta(\omega_{n},\phi)$ has $d$ and $s$ components.
The renormalization function $Z(\omega,\phi')=Z_{s}(\omega)$ has just the $s$ component because
the equation for $Z_{d}(\omega)$ is a homogeneous integral
equation whose only solution in the weak-coupling regime is
$Z_{d}(\omega)=0$ \cite{zetad}. 
The cut-off energy is $\omega_{c}=450$ meV and the maximum quasiparticle energy is $\omega_{max}=500$ meV.
In first approximation, we put $\mu^{*}_{d}=0$ (if the $s$ component of the gap is zero, the value of $\mu^{*}_{s}$ is irrelevant). If the $s$ component of the gap function is absent, this means that
$\mu^{*}_{s} \gg \mu^{*}_{d}\simeq0$ \cite{Dwave4}
\begin{figure}[ht]
    \centering
    \includegraphics[width=0.9\linewidth]{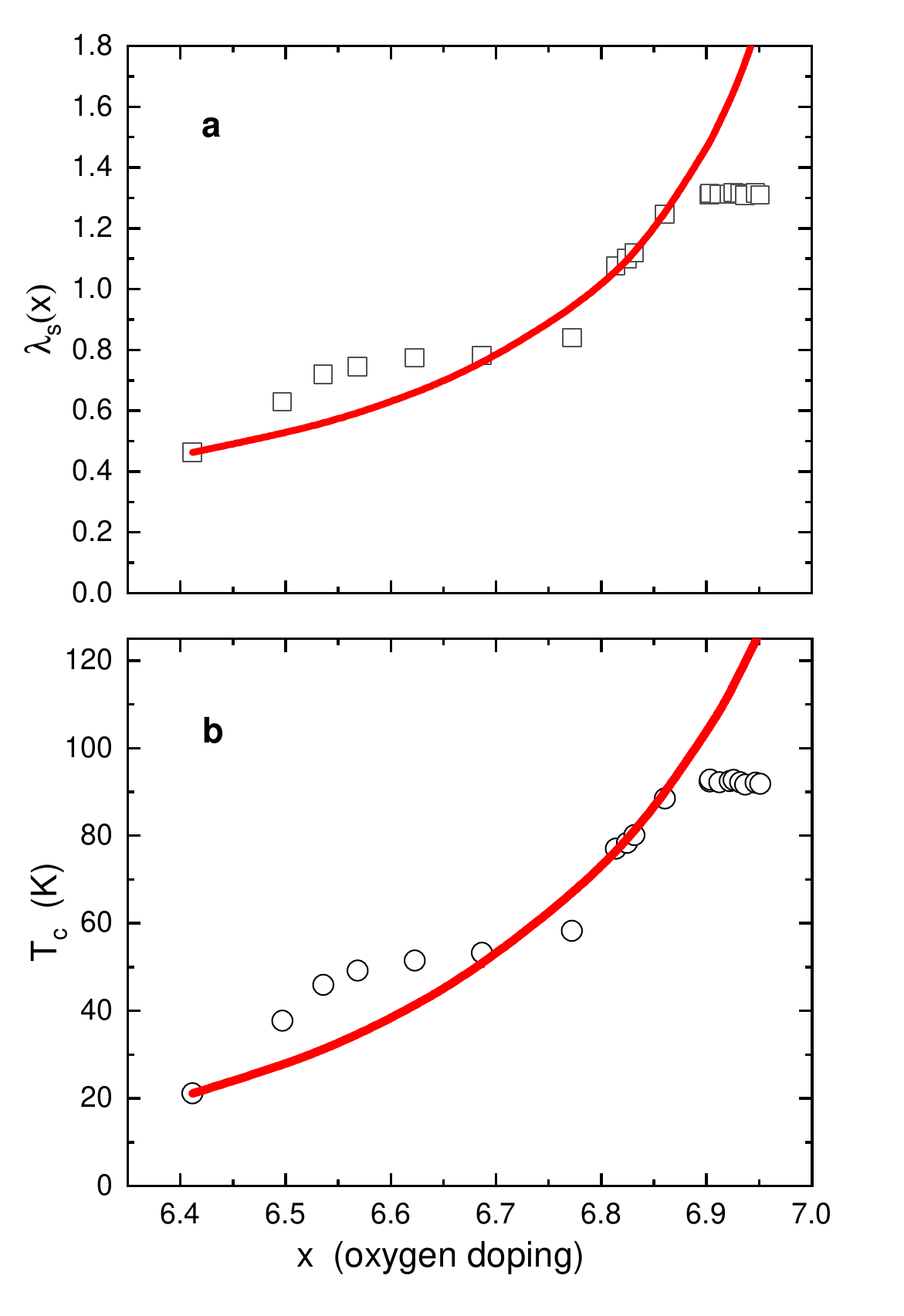}
    \caption{\textbf{(a)}: The value of the electron-phonon coupling constant $\lambda_{s}(x)$ used in the theoretical model for different values of the doping level $x$ that reproduce exactly the experimental critical temperature, is shown (open black squares). \textbf{(b)}: The experimental critical temperatures, for different values of the doping level $x$, are shown (open black squares). The continuous red line is the calculated critical temperature versus x obtained by solving the d-wave Eliashberg equations.}
    \label{fig:5}
\end{figure}
In this model, the experimental input parameters are $\alpha^{2}F_{s,d}(\Omega)$ and $\mu^{*}_{s,d}$ but, with our previous assumptions, we have just one free parameter: $\lambda_{s}(x)$.
To start with, we solve the imaginary axis d-wave Eliashberg equations and we fix the value of $\lambda_{s}(x)$ in such a way to obtain the exact experimental critical temperature $T_c$. The corresponding obtained values of $\lambda_{s}(x)$ are shown in Fig. \ref{fig:5}\color{red}(a)\color{black}. We can see that the value of the coupling constant increases by a factor three over the examined oxygen doping $x$ range, but remains in a moderate strong-coupling regime, \textit{i.e.} less than lead (Pb), for example ($\lambda_{Pb}=1.55$). The fact that the coupling constant increases by a factor three is connected with the modest variation of the representative energy of the system, which actually decreases as $x$ increases. As we know from theory, $T_{c}$ is an increasing function of the representative energy of the phonon spectrum (in our case, roughly, $\Omega_{0}$, the value of the peak of the phonon Lorentzian spectrum) and of the electron-phonon coupling constant.

At this point we recall that $\lambda_{s}(x) \propto \frac{1}{N(E_{F},x)}$ and we use the first value of $\lambda_{s}(x=6.41)=0.463$ that produces the exact experimental $T_{c}(x=6.41)=21.2$ $K$ for calculating the other values according to the following law: $\lambda_{s}(x)=\frac{N(E_{F},x=6.41)}{N(E_{F},x)}\lambda_{s}(x=6.41)\simeq [1+\frac{1}{N(E_{F},x=6.41)}\frac{\partial N(E_{F},x=6.41)}{\partial x}(x-6.41)]\lambda_{s}(x=6.41)=[1+k(x-6.41)]\lambda_{s}(x=6.41)$ where $k=\frac{1}{N(E_{F},x=6.41)}\frac{\partial N(E_{F},x=6.41)}{\partial x}$. The fit is shown in Fig. \ref{fig:5}\color{red}(a)\color{black}. We find $k=-1/4$. Now we can calculate the critical temperature without free parameters. The final results are shown in Fig. \ref{fig:5}\color{red}(b)\color{black}. Our model is too simple to explain the plateau of $T_{c}$ for $x$ close to $7$, but in its extreme simplicity it manages to account for the trend of $T_{c}$ in almost the entire oxygen doping range. Studies on the dependence of the density of states at the Fermi level, as a function of oxygen doping, would be needed to verify our hypothesis in greater detail. 

A qualitative argument to explain the decreasing trend of $N(E_F)$ vs $x$ goes as follows. Assuming an electronic DOS for the free electrons, $N(E) \sim E^{1/2}$, at the Fermi level we have $N(E_F) \sim E_F^{1/2}$. For the free electrons, $\sqrt{E_F} \sim n^{1/3}$. At chemical equilibrium between electrons and holes, $n\,p = const$, hence $n \sim 1/p$. Hence, upon further taking $p \sim x$ \cite{Talantsev}, we get $N(E_F) \sim x^{-1/3}$. Hence, even within the simplest assumptions, it is physically meaningful that the DOS at Fermi level decreases with increasing the level of doping $x$.

\section{Conclusions}
In conclusion, we have developed a simple theoretical framework to rationalize the correlation between optical phonon softening and superconducting $T_c$ in YBa$_2$Cu$_3$O$_x$, in excellent qualitative agreement with experimental data. The theoretical model is based on the Migdal-Eliashberg theory of strong-coupling superconductivity with the pairing being mediated by the $A_g$ oxygen mode that undergoes strong softening upon increasing the doping $x$ as observed experimentally \cite{Liarokapis1} and couples strongly to the nodal electronic states \cite{Devereaux2004}. 

We presented two versions of this model.
In the first version, we used a simplified Eliashberg-type argument involving three free parameters. One is the effective electron-phonon matrix element $C$, the other is the width $\beta(x)$ of the softened optical phonon of the in-phase $A_g$ oxygen Raman-active mode, and the last one is the location of the softening region within the Brillouin zone, $q_{\text{soft}}$. We emphasize that, out of these three parameters, only two, the width of the softening region in the phonon dispersion, and the parameter $C(x)$, depend on the doping level $x$. In particular, we found that the experimental data can be reproduced if $C(x)$ grows linearly with $x$. A physical justification is provided in terms of the Bardeen-Pines electron-phonon matrix element, which is approximately inversely proportional to the electron concentration $n$. Assuming chemical equilibrium between electron and holes, and since the hole concentration $p$ is proportional to the oxygen doping $x$, we showed that indeed $C(x)$ increases linearly with $x$.
The analysis using this model reveals a connection between the doping and charge order-induced softening of the in-phase oxygen vibrational mode and the superconducting $T_c$, which is otherwise difficult, if not impossible, to explain with alternative theoretical approaches (\textit{e.g.}, spin fluctuations \cite{Scalapino1999}). 
The proposed model provides a link between the two phenomena and suggests a moderate role of the phonon linewidth $\gamma(x)$ in such a regime of doping. Additionally, the model predicts that the additional plateau feature observed at optimally doping, $x>6.9$, is a direct manifestation of the doping dependence of the softening width $\beta(x)$, which cannot be neglected anymore for optimally doping levels. In other words, the extra plateau in $T_c$ vs doping $x$ past the flattening region, in the optimally doped range $x>6.9$, may be ascribed to the strong sharpening of the softening dip of the oxygen phonon mode in that range. An otherwise constant-with-doping softening width would lead to an unphysical monotonically growing $T_c$ as a function of $x$, which is not observed experimentally. Interestingly, the theoretical model returns, as one of its output, the momentum-resolved dispersion relation of the in-phase oxygen mode with a width of the softening that decreases with the depth of the softening, in qualitative agreement with typical observations of optical phonon softening in superconductors.

The second version of the model is based on the same input in terms of $A_g$ phonon mode softening, from the experimental data, which is implemented in the d-wave Eliashberg theory formalism. This is a more appropriate and exact method, which does not rely on the Allen-Dynes formula (which works, in an approximate way, just for intermediate coupling, while it fails for the strong coupling regime).
This second version of the model by calibrating the doping-dependence of the electron-phonon coupling $\lambda$ on a single value of oxygen doping, is able to reproduce the $T_c$ values as a function of oxygen doping level $x$ in excellent agreement with the experimental data. The calibration implies that the density of states (DOS) of electrons at the Fermi level is a monotonic decreasing function of $x$, which can be rationalized, once again, using a simple argument which assumes a Fermi DOS and the chemical equilibrium between electrons and holes.

The present model provides a simple solution to a few puzzles in linking doping-induced anomalous softening of oxygen modes and superconducting critical temperature in the cuprates and may serve as a useful tool for engineering and discovering new high-temperature materials in future work in conjunction with atomistic approaches \cite{EPW}. Our findings also emphasize that the role of lattice distortions in high-$T_c$ superconductors might be more important than currently thought and should probably be carefully revisited.

Finally, the role of charge order has been simplified in our model. The important role of charge order leading to anomalous softening and suppression of $T_c$ has been shown by several studies \cite{cdw1,cdw2,Yu2021}. Hence, further work on the interplay between phonon, charge order, and superconductivity, is needed.

\appendix
\section{Low-temperature Raman data}
In this Appendix we show, in Fig.\ref{newfig} below, the Raman scattering data of the in-phase oxygen mode measured at liquid nitrogen temperatures ($T=77$K) in YBa$_2$Cu$_3$O$_x$. As explained in the main article, these data have intrinsically lower statistics compared to the room temperature data.

\begin{figure}[ht]
    \centering
    \includegraphics[width=\linewidth]{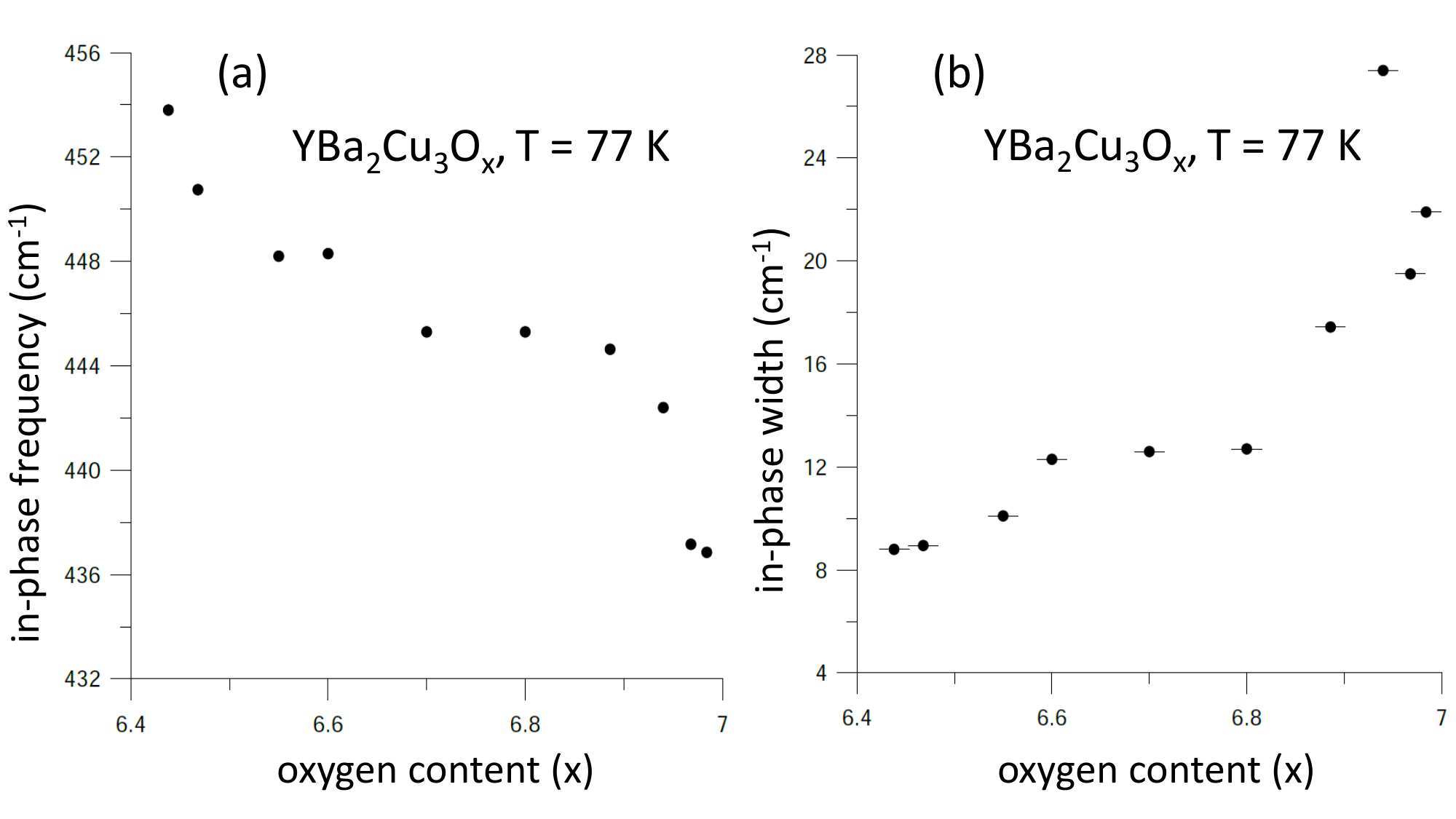}
    \caption{Experimental data of Raman scattering for the in-phase oxygen optical mode taken at $T=77$K. \textbf{(a)} shows the phonon frequency as a function of the oxygen doping content $x$. \textbf{(b)} shows the phonon linewidth as a function of the oxygen doping content $x$. }
    \label{newfig}
\end{figure}

Nevertheless, these data clearly show the same characteristics of in-phase oxygen optical phonon softening that are retained in the Raman scattering measurements at room temperature (Fig. 1). The latter were used as input to our theoretical model in the main article.

\section*{Aknowledgments}
We would like to thank Chandan Setty for collaboration on related topics and useful comments on a preliminary version of this draft.
C.J. and M.B. acknowledge the support of the Shanghai Municipal Science and Technology Major Project (Grant No.2019SHZDZX01) and the sponsorship from the Yangyang Development Fund. A.Z. gratefully acknowledges funding from the European Union through Horizon Europe ERC Consolidator Grant number: 101043968 ``Multimech'', and from US Army Research Office through contract nr.W911NF-22-2-0256.

\section*{Data availability}
The data that supports the findings of this study are available within the article.

\bibliographystyle{apsrev4-1}
\bibliography{ref}

\end{document}